\documentclass[aps,prl,twocolumn,floatfix,letterpaper,longbibliography]{revtex4-2}
\usepackage{graphicx}
\usepackage{bm}
\usepackage{amsmath,amsfonts}
\usepackage{amsbsy}
\usepackage[colorlinks=true,linkcolor=blue,urlcolor=blue,citecolor=blue,breaklinks=true]{hyperref}
\usepackage[utf8]{inputenc}
\DeclareUnicodeCharacter{03BC}{\mbox{$\mu$}}
\DeclareUnicodeCharacter{2009}{ }

\def\TITLE{Spectral function of chiral one-dimensional Fermi liquid
  in the regime of strong interactions}

\begin{document}

\title{\TITLE}

\author{K. A. Matveev}

\affiliation{Materials Science Division, Argonne National Laboratory,
  Argonne, Illinois 60439, USA}

\date{March 2, 2022}

\begin{abstract}

  We study momentum-resolved tunneling into a system of spinless
  chiral one-dimensional fermions, such as electrons at the edge of an
  integer quantum Hall system.  Interactions between particles give
  rise to broadening of the spectral function of the system.  We
  develop an approach that enables one to obtain the shape of the peak
  in the spectral function in the regime of strong interaction.  We
  apply our technique to the special cases of short-range and Coulomb
  interactions.

\end{abstract}
\maketitle

Low energy properties of one-dimensional systems of fermions are
strongly affected by interactions between them.  These systems are
commonly described in terms of the Luttinger liquid theory
\cite{haldane_luttinger_1981, giamarchi_quantum_2004}.  The signature
feature of the Luttinger liquid is the power-law behavior of the
tunneling density of states at low energies
\cite{kane_transmission_1992, furusaki_single-barrier_1993}.  A
special kind of one-dimensional system---the chiral Luttinger
liquid---has been predicted to emerge at the edge of the fractional
quantum Hall system \cite{wen_theory_1992}.  Its properties are
controlled by the nature of the quantum Hall state; specifically, its
occupation fraction $\nu$.  For systems where $\nu^{-1}$ is odd
integer, the tunneling density of states scales as a power of energy,
$D(\epsilon)\propto\epsilon^\alpha$, with the exponent
$\alpha=\nu^{-1}-1$ \cite{wen_theory_1992,chang_chiral_2003}.
Importantly, even though the interactions are crucial for the
formation of the fractional quantum Hall state, their strength does
not affect the value of the exponent $\alpha$.

In this paper we study the model of spinless chiral one-dimensional
fermions, which can be realized experimentally as the edge mode of an
integer quantum Hall system with occupation fraction $\nu=1$
\cite{halperin_quantized_1982}.  The above results can be applied to
this special case and yield the exponent $\alpha=0$, i.e., the density
of states at the Fermi level assumes a constant value.  This
conclusion applies regardless of the strength of interaction between
the fermions, and in this respect the system behaves as a Fermi rather
than Luttinger liquid.  The latter statement does not imply that the
elementary excitations of the system are fermions.  Their nature
depends on momentum and interaction strength.  In particular, in the
case of Coulomb repulsion the low energy excitations are bosons
\cite{2109.06220}.

Although interactions between fermions in this model do not lead to
power-law scaling of the tunneling density of states, they manifest
themselves in the spectral function of the system $A_p(\epsilon)$.
The latter describes momentum-resolved tunneling into the system and
provides more detailed information than the density of states
$D(\epsilon)$.  In the absence of interaction between particles, the
spectral function has a sharp peak at the energy $\epsilon_p$ of the
fermion of momentum $p$, i.e.,
$A_p(\epsilon)=\delta(\epsilon-\epsilon_p)$.  Interactions result in
broadening of this peak.  Our particular focus will be on the limit of
strong interactions, where the elementary excitations with momenta of
order $p$ are bosonic.  Our main result is the exact shape of the peak
in the spectral function in this regime.

% Although the bosonic nature of the elementary excitations in this
% model does not lead to power-law scaling of the tunneling density of
% states, it manifests itself in the spectral function of the system
% $A_p(\epsilon)$.  The latter describes momentum-resolved tunneling
% into the system and provides more detailed information than the
% density of states $D(\epsilon)$.  In the absence of interaction
% between particles, the spectral function has a sharp peak at the
% energy $\epsilon_p$ of the fermion of momentum $p$, i.e.,
% $A_p(\epsilon)=\delta(\epsilon-\epsilon_p)$.  Interactions result
% in broadening of this peak.  Our particular focus will be on the limit
% of strong interactions, where the elementary excitations with momenta
% of order $p$ are bosonic.  Our main result is the exact shape of the
% peak in the spectral function in this regime.

Our approach to the evaluation of $A_p(\epsilon)$ is complementary to
that developed for non-chiral systems in
Ref.~\cite{imambekov_phenomenology_2009}.  The latter applies at any
interaction strength, but is limited to the evaluation of the spectral
function only at the edge of support, where $A_p(\epsilon)$ exhibits
power-law scaling.  Our approach gives $A_p(\epsilon)$ for all
energies, but is limited to the regime of strong interactions.

It is worth pointing out that the chiral nature of our problem ensures
that for short-range interaction the spectral function vanishes
outside of a finite interval of energies.  Assuming that interactions
are repulsive, the lower edge of support is analogous to that
considered for the non-chiral case in
Ref.~\cite{imambekov_phenomenology_2009}, and there the results can be
compared.  On the other hand, the upper edge of support of
$A_p(\epsilon)$ is qualitatively different and cannot be treated using
the method of Ref.~\cite{imambekov_phenomenology_2009}.  The fermion
tunneling into the system near that energy generates a very large
number of bosonic excitations.  The corresponding physics is similar
to that of the multiphonon spectrum of liquid helium
\cite{iordanskii_properties_1978} and multiparticle production in
quantum field theory \cite{khoze_review_2019}.  Finally, our approach
is not limited to the case of short-range interactions.  As an
example, we also consider the case of Coulomb interactions, whose
long-range nature changes the spectral function qualitatively.

% Tunneling of a fermion into the system changes its momentum by $p$. In
% a chiral system with short-range interactions, the resulting energy
% can only take values in a finite band, beyond which the spectral
% function vanishes.  Assuming strong repulsive interactions, the lower
% edge of this band corresponds to exciting a single bosonic excitation
% with momentum $p$.  Behavior of the spectral function near this edge
% can be compared with the results for the non-chiral system
% \cite{imambekov_phenomenology_2009}.  For energies near the upper edge
% of support of $A_p(\epsilon)$, the momentum $p$ is distributed among a
% large number of bosonic excitations with the total momentum $p$.  This
% results in exponential suppression of the spectral function near the
% upper edge.  The corresponding physics is analogous that of the
% multiphonon spectrum of liquid helium
% \cite{iordanskii_properties_1978} and multiparticle production in
% quantum field theory \cite{khoze_review_2019}.

We describe our system by the Hamiltonian $\hat H=\hat
H_0+\hat V$, where
\begin{eqnarray}
  \label{eq:H_0}
  \hat H_0&=&\sum_p\epsilon_p c_p^\dagger c_p^{},
  \quad
  \epsilon_p=vp+\frac{p^2}{2m}+\ldots,
  \\
  \label{eq:V}
  \hat V&=&\frac{1}{L}
  \sum_{\substack{p_1 p_2\\ q>0}}
  V(q)c_{p_1+q}^\dagger c_{p_1}^{} c_{p_2}^\dagger
      c_{p_2+q}^{}.
\end{eqnarray}
Here $c_p$ is the fermion annihilation operator.  The energy of the
fermion $\epsilon_p$ is assumed to be a monotonic function of momentum
$p$, which ensures that the fermions are chiral.  The momentum and
energy are measured from their values at the Fermi point.  The system
has a finite length $L$; we assume periodic boundary conditions.
$V(q)$ is the Fourier transform of the interaction potential.

Our main interest is the spectral function of the system defined as
\begin{equation}
  \label{eq:SFdefinition}
  A_p(\epsilon)=\sum_j|\langle j|c_p^\dagger|0\rangle|^2
  \delta(\epsilon-E_j).
\end{equation}
Here $|0\rangle$ denotes the eigenstate of $\hat H$ in which all
single-particle states with $p<0$ are filled and those with $p\geq0$
empty.  The summation is over all the eigenstates $|j\rangle$ of the
Hamiltonian $\hat H$, and $E_j$ is the energy of the state $|j\rangle$
measured from that of the vacuum state $|0\rangle$.

In the absence of interactions, the sum in Eq.~(\ref{eq:SFdefinition})
contains only one nonvanishing term, which corresponds to
$|j\rangle=c_p^\dagger|0\rangle$.  In this case we immediately obtain
$A_p(\epsilon)=\delta(\epsilon-\epsilon_p)$.  In an interacting
system, multiple eigenstates of the Hamiltonian
$\hat H=\hat H_0+\hat V$ overlap with
$c_p^\dagger|0\rangle$, and the spectral function $A_p(\epsilon)$ has
the meaning of the distribution function of the energy of the system
after a single fermion with momentum $p$ is added to it.

The width of the energy distribution can be quantified by evaluating
its variance
$\delta\epsilon^2=\overline{\epsilon^2}-\overline\epsilon^2$, where
the averaging is performed by integrating with the weight
$A_p(\epsilon)$.  Using the definition (\ref{eq:SFdefinition}), it can
be expressed as
\begin{equation}
  \label{eq:variance}
  \delta\epsilon^2=\langle0|c_p \hat H^2c_p^\dagger|0\rangle
  -\langle0|c_p \hat H c_p^\dagger|0\rangle^2.
\end{equation}
Substituting $\hat H=\hat H_0+\hat V$ given by Eqs.~(\ref{eq:H_0}) and
(\ref{eq:V}), in the thermodynamic limit $L\to\infty$ we find
% \begin{equation}
%   \label{eq:variance_result}
%   \delta\epsilon^2=\frac{1}{(2\pi\hbar)^2}
%   \int_0^p\!dq\int_0^q\!dk [V(q)-V(p-k)]^2\theta(p-q-k),
% \end{equation}
%where $\theta(x)$ is the unit step function.
\begin{equation}
  \label{eq:variance_result}
  \delta\epsilon^2=\frac{1}{8\pi^2\hbar^2}
  \int_0^pdq\int_{p-q}^pdq' [V(q)-V(q')]^2.
\end{equation}

To illustrate our results, we will apply them to two types of
interaction potential.  First, we will consider short-range
interactions that fall off with the distance fast enough to ensure
that $V(q)$ and its second derivative $V''(q)$ are well defined at
$q=0$.  In this case we will approximate
\begin{equation}
  \label{eq:Vshort}
  V(q)=V(0)-\eta q^2,
  \quad
  \eta=-\frac12 V''(0).
\end{equation}
For simplicity, we will assume $\eta>0$, which corresponds to a
typical repulsive interaction.  Substituting this expression into
Eq.~(\ref{eq:variance_result}), we obtain
\begin{equation}
  \label{eq:variance_short}
  \delta\epsilon^2=\frac{11}{720\pi^2\hbar^2}\eta^2p^6.
\end{equation}
In addition, we will consider Coulomb interaction $e^2/|x|$ cut off at
a short distance $w$.  In this case we have
\begin{equation}
  \label{eq:VCoulomb}
  V(q)=2e^2\ln\frac{\hbar}{|q|w}
\end{equation}
and
\begin{equation}
  \label{eq:variance_Coulomb}
  \delta\epsilon^2= \left(\frac{1}{\pi^2}-\frac{1}{12}\right)
         \frac{e^4p^2}{\hbar^2}.
\end{equation}
It is worth mentioning that the expression (\ref{eq:variance_result})
for the variance of the energy distribution and the subsequent results
(\ref{eq:variance_short}) and (\ref{eq:variance_Coulomb}) are not
pertubative in $\hat V$; they apply for any interaction strength.

Let us now discuss the effect of different parts of the Hamiltonian
$\hat H$ on the spectral function.  At small $p$ the largest
contribution is $\sum_pvpc_p^\dagger c_p^{}$, obtained from
Eq.~(\ref{eq:H_0}) by linearizing the energy spectrum,
$\epsilon_p=vp$.  This term is simply $v\hat P$, where $\hat P$ is the
momentum operator.  In our system, momentum is conserved, and $v\hat
P$ commutes with the remaining terms of the Hamiltonian.  Its effect
on the spectral function (\ref{eq:SFdefinition}) amounts to adding
$vp$ to the energy of each state $|j\rangle$, and thus to the shift
$vp$ of the peak in $A_p(\epsilon)$ as a function of energy at a given
$p$.  Importantly, this term does not affect the shape of the peak.

The leading term describing the curvature of the spectrum $\epsilon_p$
in Eq.~(\ref{eq:H_0}) is
$\delta \hat H_0=\sum_p(p^2/2m)c_p^\dagger c_p^{}$.  It does not
commute with the interaction term $\hat V$.  As a result, the
evaluation of the spectral function in the regime where
$\delta\hat H_0$ and $\hat V$ have comparable effects on it, is a
challenging problem.  To make further progress, we determine which of
the two contributions to the Hamiltonian, $\delta\hat H_0$ or
$\hat V$, is dominant at a given $p$ and interaction strength.

If the interactions are neglected, the effect of $\delta \hat H_0$ on
the spectral function amounts to the additional shift of the peak by
$\delta \epsilon_p = p^2/2m$.  On the other hand, the interaction
$\hat V$ results in broadening of the peak, with the characteristic
width $\delta\epsilon$ given by Eq.~(\ref{eq:variance_result}).
Comparison of $\delta\epsilon_p$ and $\delta\epsilon$ indicates which
of the two perturbations gives the dominant contribution to
$A_p(\epsilon)$.  At $\delta\epsilon_p\gg\delta\epsilon$ the effect of
interactions is small, and $\hat V$ can be accounted for in
perturbation theory \cite{khodas_fermi-luttinger_2007}.  Below we
study in detail the opposite case,
$\delta\epsilon\gg\delta\epsilon_p$, in which the shape of the
spectral function $A_p(\epsilon)$ is controlled by interactions.  For
short-range interactions, using Eq.~(\ref{eq:variance_short}) we find
that the interaction dominated regime is achieved at
$m\eta p/\hbar\gg1$.  For the Coulomb interaction,
Eq.~(\ref{eq:variance_Coulomb}) yields the condition
$me^2/\hbar p\gg1$.  Note that while for the short-range interactions
the condition requires relatively high momentum $p\gg\hbar/m\eta$,
in the Coulomb case interactions dominate at small momentum,
$p\ll me^2/\hbar$.

The treatment of the problem in the limit of strong interactions is
greatly simplified by bosonizing fermion operators.  In the bosonic
representation the state of the system is described by occupation
numbers of bosonic states numbered by $l=1,2,\ldots$ and the total
number of fermions $N$, which can be measured from that in the ground
state $|0\rangle$.  The fermion field operator is given by
\cite{haldane_luttinger_1981}
\begin{equation}
  \label{eq:bosonization}
  \Psi(x)=\frac{1}{\sqrt L}
  U e^{ik_F x}e^{i\varphi^\dagger(x)}e^{i\varphi(x)}.
\end{equation}
Here $k_F$ is the Fermi wavevector for a given number of particles
$N$, the operator $U$ lowers $N$ by 1, and $\varphi(x)$ is related to the
bosonic destruction operators $b_l$ by
\begin{equation}
  \label{eq:phi}
  \varphi(x)=-i\sum_{l=1}^\infty
  \frac{1}{\sqrt l}e^{iq_l^{} x/\hbar}b_l,
  \quad
  q_l^{}=\frac{2\pi\hbar}{L}l.
\end{equation}
The advantage of using bosonization is that for fermions with linear
spectrum, $\epsilon_p=vp$, the
Hamiltonian (\ref{eq:H_0}), (\ref{eq:V}) takes the simple form
\begin{equation}
  \label{eq:Hb}
  \hat H_b =
  \sum_l \varepsilon_l^{} b_l^\dagger b_l^{},
  \qquad
  \varepsilon_l^{}=\left(v+\frac{V(q_l^{})}{2\pi\hbar}\right)q_l^{}.
\end{equation}
Importantly, $\hat H_b$ is quadratic in the bosonic operators, i.e.,
it describes a system of noninteracting bosonic elementary excitations
with energies $\varepsilon_l^{}$.  The effects of the curvature of the
spectrum would generate interactions of the bosons.  From now on we
limit ourselves to the strong interaction limit and thus neglect these
interactions.

We start by rewriting the definition (\ref{eq:SFdefinition}) of the
spectral function in terms of the fermionic Green's function,
\begin{equation}
  \label{eq:SFAltDefinition}
  A_{p_k}(\epsilon)
  =\int_{-\infty}^\infty \frac{dt}{2\pi\hbar}
  \int_0^L dx\, e^{-i(p_k x-\epsilon t)/\hbar}
  g(x,t),
\end{equation}
where $g(x,t)=\langle0| \Psi(x,t)\Psi^\dagger(0,0)|0\rangle$.  To
account for the finite size $L$ of the system, we replaced
$p\to p_k=2\pi\hbar k/L$, where $k$ is integer.  Given the simple
quadratic form of the Hamiltonian (\ref{eq:Hb}) and the bosonized form
(\ref{eq:bosonization}) of the operator $\Psi$, the evaluation of the
Green's function is straightforward,
\begin{equation}
  \label{eq:greens}
  g(x,t)
  =\frac{1}{L}
  \exp\left(\sum_{l=1}^\infty \frac{1}{l }
  e^{i(q_l^{}x-\varepsilon_l^{} t)/\hbar}\right).
\end{equation}

Expansion of the exponential in Eq.~(\ref{eq:greens}) in Taylor series
upon substitution into Eq.~(\ref{eq:SFAltDefinition}) immediately
yields
\begin{eqnarray}
  \label{eq:SFNegative-p}
  A_{p_k}(\epsilon)&=&0,
  \quad k<0,
                       \\
  \label{eq:SF-k=0}
  A_{p_0}(\epsilon)&=&\delta(\epsilon),
                       \\
  \label{eq:SF-general-k}
  A_{p_k}(\varepsilon)&=&\sum_{\{N_l\}_k}
  \!\delta\Bigg(\epsilon-\sum_{l=1}^\infty N_l\varepsilon_l^{}\Bigg)
  \!\prod_{l=1}^k\frac{1}{N_l!\,l^{N_l}},
  \quad
  k>0.\hspace{2em}
\end{eqnarray}
Equation (\ref{eq:SFNegative-p}) is expected because in $|0\rangle$
all the fermion states with negative momenta are filled. The single
$\delta$-function on the right-hand side of Eq.~(\ref{eq:SF-k=0})
accounts for the only possible state of the system with extra particle
and momentum $0$.  For $k>0$ the state $c_{p_k}^\dagger|0\rangle$ is a
superposition of all possible states involving bosons with
$l=1,2,\ldots,k$ and occupation numbers $N_l$, such that
$\sum_lN_ll=k$, as required by momentum conservation.  Thus, the
summation in Eq.~(\ref{eq:SF-general-k}) is over all integer
partitions of $k$.

As $k$ increases, the number of terms in Eq.~(\ref{eq:SF-general-k}),
which is the number of integer partitions of $k$, grows exponentially
\cite{hardy_asymptotic_1918}.  Physically, we are interested in the
thermodynamic limit, whereby $L\to\infty$ and $k\to\infty$ so that
$p_k=2\pi\hbar k/L$ takes a fixed value $p$.  It is not immediately
clear how to take this limit in Eq.~(\ref{eq:SF-general-k}).  Instead,
we return to Eq.~(\ref{eq:SFAltDefinition}), replace
$e^{-ip_k x/\hbar}\to(i\hbar/p_k)\partial_xe^{-ip_k x/\hbar}$, and
integrate by parts with respect to $x$.  Using Eq.~(\ref{eq:greens})
we then find
\begin{equation}
  \label{eq:recurrence}
  A_{p_k}(\epsilon)=\frac{1}{k}\sum_{l=1}^k A_{p_{k-l}^{}}(\epsilon-\varepsilon_l).
\end{equation}
Here we limited the summation to $l\leq k$ because $A_{p_{k-l}}=0$ at
$l>k$, see Eq.~(\ref{eq:SFNegative-p}).  Equation
(\ref{eq:recurrence}) is a recurrence relation that enables one to
obtain the spectral function for any positive $k$, given known
expressions for $k'=0,1,\ldots,k-1$.  In the thermodynamic limit we
convert the sum in Eq.~(\ref{eq:recurrence}) into an integral and find
\begin{equation}
  \label{eq:integral-equation}
  A_p(\epsilon)=\frac{1}{p}\int_0^p dq
  A_{p-q}(\epsilon-\varepsilon(q)),
  \quad
  \varepsilon(q)=\left(v+\frac{V(q)}{2\pi\hbar}\right)q.
\end{equation}
For a given interaction $V(q)$, the spectral function can be found by
solving the integral equation (\ref{eq:integral-equation}).

We now use Eq.~(\ref{eq:integral-equation}) to evaluate
$A_p(\epsilon)$ in the case of short-range interactions.  Using
Eq.~(\ref{eq:Vshort}), we get
\begin{equation}
  \label{eq:spectrum-short}
  \varepsilon(q)=\tilde v q -\frac{\eta}{2\pi\hbar}q^3,
  \qquad
  q>0,
\end{equation}
where $\tilde v=v+V(0)/2\pi\hbar$.  Because the function
$\varepsilon(q)$ is concave, the total energy $\epsilon$ of any set of
bosonic excitations with the total momentum $p$ is in the range
$\varepsilon(p)\leq\epsilon<\tilde vp$.  Thus, it is convenient to
look for a solution of the integral equation
(\ref{eq:integral-equation}) in the form
\begin{equation}
  \label{eq:parametrization-short}
  A_p(\epsilon)=\frac{2\pi\hbar}{\eta p^3}
  a\left(
    \frac{2\pi\hbar}{\eta p^3}
    [\epsilon-\varepsilon(p)]
  \right),
\end{equation}
where $a(\lambda)$ is a dimensionless function that vanishes outside
the interval $0\leq\lambda<1$ and satisfies the normalization
condition $\int_0^1 a(\lambda)d\lambda=1$.  Substitution of
Eqs.~(\ref{eq:spectrum-short}) and (\ref{eq:parametrization-short}) into
Eq.~(\ref{eq:integral-equation}) yields
\begin{equation}
  \label{eq:integral-equation-short}
  a(\lambda)=\int_0^1\frac{d\zeta}{\zeta^3}
  a\left(\frac{\lambda-3\zeta(1-\zeta)}{\zeta^3}\right),
\end{equation}
where introduced the integration variable $\zeta=(p-q)/p$.

\begin{figure}[t]
\includegraphics[width=.47\textwidth]{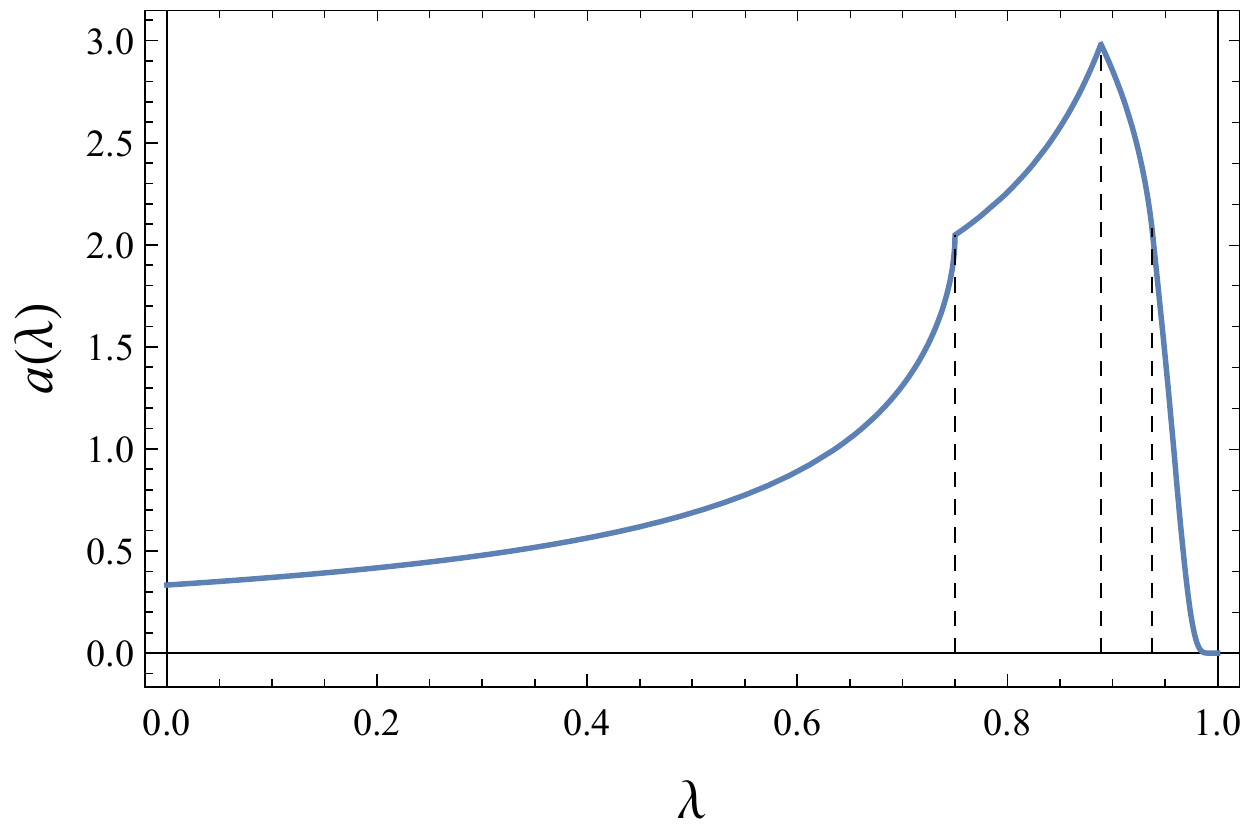}
\caption{The dimensionless spectral function $a(\lambda)$ in the case of
  short-range interactions, obtained by solving the integral equation
  (\ref{eq:integral-equation-short}) numerically.  The dashed lines
  are at $\lambda=\frac34,\frac89,\frac{15}{16}$.}
\label{fig:a_lambda_short}
\end{figure}

A numerical solution of the integral equation
(\ref{eq:integral-equation-short}) is shown in
Fig.~\ref{fig:a_lambda_short}.  A careful examination of
Eq.~(\ref{eq:integral-equation-short}) shows \cite{onefootnote} that
$a(0^+)= 1/3$, whereas at $\lambda\to1$ the solution is suppressed
exponentially,
\begin{equation}
  \label{eq:a-near-1}
  \ln a(\lambda)\simeq-\frac{1}{2\sqrt{1-\lambda}}
    \ln\frac{1}{1-\lambda},
  \quad
  \lambda\to1^-.
\end{equation}
Near the lower edge of support the spectral function is expected to
show power-law scaling with the exponent $\mu$, which depends on the
interaction strength \cite{imambekov_phenomenology_2009}.  For the
chiral system the result in the limit of strong interaction is $\mu=0$
\cite{onefootnote}, which is consistent with our conclusion that
$a(\lambda)$ approaches a finite value at $\lambda\to0^+$.  In order
for the state with momentum $p$ to have energy close to $\tilde v p$,
corresponding to $\lambda=1$, the momentum must be distributed among a
very large number of bosonic excitations.  The amplitude of such
processes is known \cite{iordanskii_properties_1978,
  khoze_review_2019} to be exponentially suppressed in a way
consistent with Eq.~(\ref{eq:a-near-1}).

\begin{figure}[t]
\includegraphics[width=.47\textwidth]{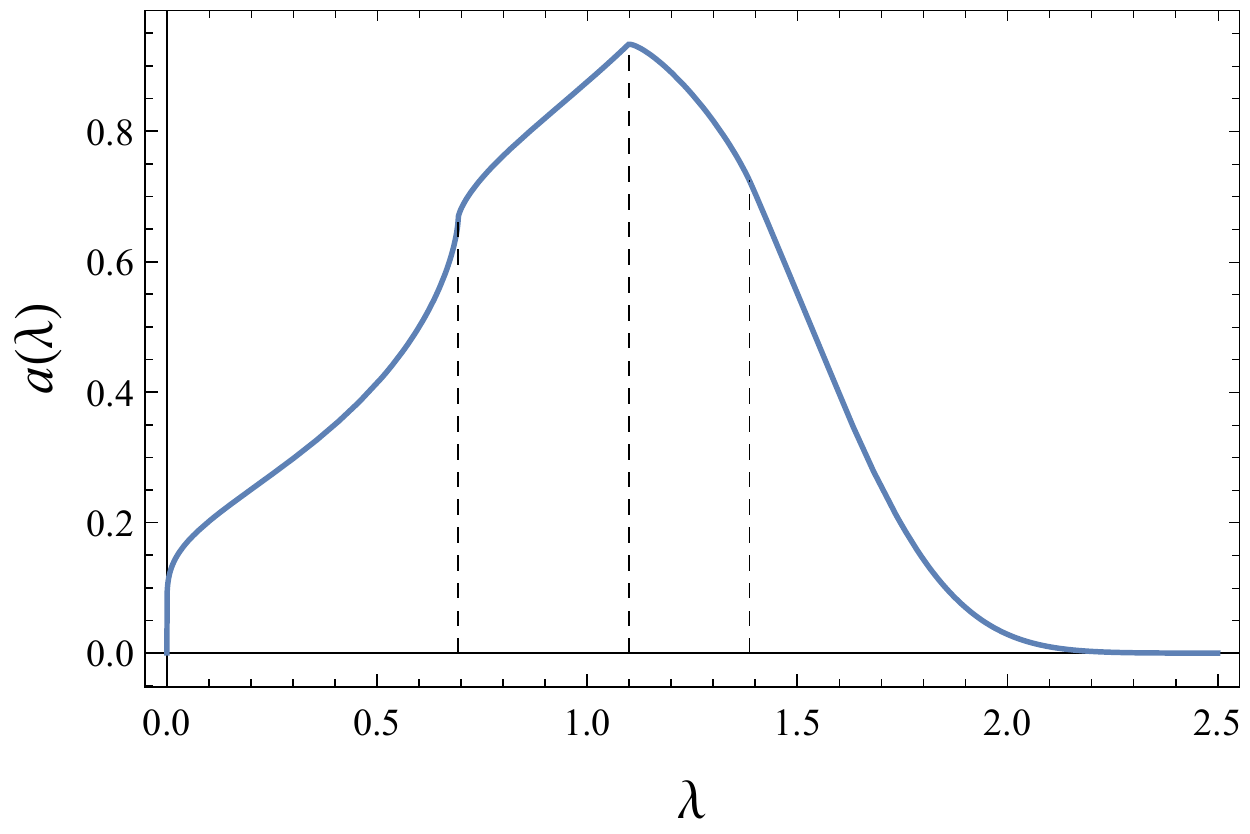}
\caption{The dimensionless spectral function $a(\lambda)$ in the case
  of Coulomb interactions, obtained by solving the integral equation
  (\ref{eq:integral-equation-Coulomb}) numerically.   The dashed lines
  are at $\lambda=\ln2$, $\ln3$, and $\ln4$.}
\label{fig:a_lambda-Coulomb}
\end{figure}

In the case of Coulomb interaction, the energies of the bosonic
excitations, obtained from Eqs.~(\ref{eq:VCoulomb}) and
(\ref{eq:integral-equation}), are
\begin{equation}
  \label{eq:spectrum-Coulomb}
  \varepsilon(q)
  =\left(
    v+\frac{e^2}{\pi\hbar}\ln\frac{\hbar}{qw}
  \right)q,
  \qquad
  q>0.
\end{equation}
Because of the long-range nature of the interaction, the velocity of
the excitations diverges at $q\to0$.  As a result, while the energy of
any state with the total momentum $p$ is limited by $\varepsilon(p)$
from below, it is not limited from above.  The typical range of
energies can be estimated as
$2\varepsilon(p/2)-\varepsilon(p)\sim(e^2/\pi\hbar)p$. This suggests
the following rescaling of the spectral function
\begin{equation}
  \label{eq:parametrization-Coulomb}
  A_p(\epsilon)=\frac{\pi\hbar}{e^2p}
  a\left(
    \frac{\pi\hbar}{e^2p}
    [\epsilon-\varepsilon(p)]
  \right),
\end{equation}
where the new dimensionless function $a(\lambda)$ vanishes at
$\lambda<0$ and satisfies the normalization condition $\int_0^\infty
a(\lambda)d\lambda=1$.  Substitution of
Eq.~(\ref{eq:parametrization-Coulomb}) into the integral equation
(\ref{eq:integral-equation}) yields
\begin{equation}
  \label{eq:integral-equation-Coulomb}
  a(\lambda)=
  \int_0^1\frac{d\zeta}{\zeta}
  a\Bigg(
    \frac{1}{\zeta}
    \Bigg(
      \lambda
      -\zeta\ln\frac{1}{\zeta}
      -(1-\zeta)\ln\frac{1}{1-\zeta}
    \Bigg)
  \Bigg).
\end{equation}
A numerical solution of the integral equation
(\ref{eq:integral-equation-Coulomb}) is shown in
Fig.~\ref{fig:a_lambda-Coulomb}.  In contrast to short-range
interactions, the spectral function vanishes at the lower bound
\cite{onefootnote},
\begin{equation}
  \label{eq:a_small_lambda_Coulomb}
  a(\lambda)\simeq\frac{1}{\ln\frac{1}{\lambda}},
  \qquad
  \lambda\to0^+.
\end{equation}
At $\lambda\gg 1$ the solution $a(\lambda)$ is exponentially small,
with $\ln a(\lambda)\simeq -\lambda e^\lambda$ \cite{onefootnote}.

The most striking feature of the spectral functions in
Figs.~\ref{fig:a_lambda_short} and \ref{fig:a_lambda-Coulomb} is the
presence of a series of singularities shown by the dashed lines at
$\lambda=1-n^{-2}$ and $\lambda=\ln n$, respectively.  The features
with $n=2$ and $3$ are most prominent, but features with $n$ up to 6
can still be seen in the plots of first and second derivatives of
$a(\lambda)$.  In both cases these positions correspond to the
energies $\epsilon=n\varepsilon(p/n)$ of the states in which the total
momentum $p$ is distributed equally among $n$ bosonic excitations.
These features can be understood from Eq.~(\ref{eq:integral-equation})
using the fact that $A_p(\epsilon)=0$ at $\epsilon<\varepsilon(p)$.
The latter condition means that the integrand in
Eq.~(\ref{eq:integral-equation}) vanishes for
$\epsilon-\varepsilon(q)<\varepsilon(p-q)$.  Since the maximum value
of $\varepsilon(q)+\varepsilon(p-q)$ is $2\varepsilon(p/2)$, the
integration region effectively changes when $\epsilon$ crosses this
value, and the spectral function has a singularity.  To understand the
feature for $n=3$, one can iterate Eq.~(\ref{eq:integral-equation}) to
express $A_p(\epsilon)$ as a double integral of
$A_{p-q-q'}(\epsilon-\varepsilon(q)-\varepsilon(q'))$ over $q$ and
$q'$ and repeat the above argument.  Successive iterations of
Eq.~(\ref{eq:integral-equation}) demonstrate the existence of the
features at $\epsilon=n\varepsilon(p/n)$ with $n=4,5,\ldots$.
Finally, we mention that in both cases of the short-range and Coulomb
interactions, the second singularity coincides with the maximum of
$a(\lambda)$, i.e., the maximum of $A_p(\epsilon)$ is at
$\epsilon=3\varepsilon(p/3)$.

Although the exact shape of the spectral function shown in
Figs.~\ref{fig:a_lambda_short} and \ref{fig:a_lambda-Coulomb} was
obtained only for the short-range and Coulomb interactions, the above
argument does not rely on the specific form of the interaction
potential.  A similar sequence of singularities is therefore expected
for other interactions, including the experimentally relevant case of
Coulomb interaction screened by a nearby gate.  Experimental
observation of these singularities would thus show explicitly the
existence of the bosonic excitations and measure their spectrum
$\varepsilon(q)$.

To summarize, we have reduced the problem of the evaluation of the
spectral function $A_p(\epsilon)$ of a system of chiral
one-dimensional fermions in the regime of strong interactions to
solving the integral equation (\ref{eq:integral-equation}).  Although
analytical solution of Eq.~(\ref{eq:integral-equation}) is not
feasible, the numerical solution for a given interaction $V(q)$ is
straightforward, see Figs.~\ref{fig:a_lambda_short} and
\ref{fig:a_lambda-Coulomb} for the short-range and Coulomb
interactions.  The solutions show sharp features at the energies
$\epsilon$ at which the momentum of the fermion is transferred to $n$
bosonic excitations with momenta $p/n$ for $n=2,3,4,\ldots$.  For the
above two types of interactions, the position of the maximum of
$A_p(\epsilon)$, is determined exactly.

\begin{acknowledgments}

  The author is grateful to L.~I.~Glazman, I. Martin and M.~Pustilnik
  for helpful discussions.  This work was supported by the
  U.S. Department of Energy, Office of Science, Basic Energy Sciences,
  Materials Sciences and Engineering Division.
  
\end{acknowledgments}

\onecolumngrid
\newpage
\setcounter{equation}{0}
\setcounter{figure}{0}

\renewcommand{\theequation}{S\arabic{equation}}
\renewcommand{\thepage}{S\arabic{page}}
\renewcommand{\thesection}{S\arabic{section}}
\renewcommand{\thetable}{S\arabic{table}}
\renewcommand{\thefigure}{S\arabic{figure}}
% use bibnumfmt to change style at the end of the document

\renewcommand{\bibnumfmt}[1]{[{\normalfont S#1}]}

% citenumfont command adds S to all numbers
%\renewcommand{\citenumfont}[1]{S#1}

\setcounter{page}{1}

\begin{center}
	{\large\textbf{\TITLE}
		\\\vskip 5pt
		\normalsize{--Supplemental Material--}
		\\\vskip 5pt
	}
	%\end{center}
	
	%\begin{center}
        K. A. Matveev \vskip 0.5mm \textit{\small Materials Science
          Division, Argonne National Laboratory, Argonne, Illinois
          60439, USA}

\end{center}
\vskip 1.5pt
\twocolumngrid

\section{I.~~Asymptotic properties of the spectral function in the
  case of short-range interactions}

Here we use the form of the integral equation
(\ref{eq:integral-equation-short}), the
normalization condition
\begin{equation}
  \label{eq:a_normalization}
  \int_0^1a(z)dz=1,
\end{equation}
and the property that $a(\lambda)$ vanishes at $\lambda<0$ and
$\lambda>1$ to obtain the behavior of $a(\lambda)$ at the edges of
support, $\lambda=0$ and 1.

\subsection{A.~~Behavior of $a(\lambda)$ at $\lambda\to0^+$.}
\label{sec:asympt-lambda-0}

Because $a(\lambda)=0$ for negative $\lambda$, at $\lambda<3/4$ the
integral in Eq.~(\ref{eq:integral-equation-short}) can be split
into two,
\begin{eqnarray}
  \label{eq:integral_equation_short_range_split}
  a(\lambda)
  &=&\int_0^{\eta}\frac{d\zeta}{\zeta^3}
      a\left(\frac{\lambda-3\zeta(1-\zeta)}{\zeta^3}\right)
\nonumber\\
  &&+ \int_{1-\eta}^1\frac{d\zeta}{\zeta^3}
  a\left(\frac{\lambda-3\zeta(1-\zeta)}{\zeta^3}\right),
\end{eqnarray}
where
\[
  \eta(1-\eta)=\frac\lambda3,
  \quad
  \eta<\frac12.
\]
To leading order at $\lambda\to0$ we have $\eta=\lambda/3$.  Since
$a(\lambda)\sim1$, the second term in
Eq.~(\ref{eq:integral_equation_short_range_split}) is of the order of
$\eta\sim\lambda$.  We will see below that this is smaller than the
contribution of the first term, i.e., to leading order the behavior of
$a(\lambda)$ at $\lambda\to0$ can be obtained from
\[
  a(\lambda)
  =\int_0^{\eta}\frac{d\zeta}{\zeta^3}
      a\left(\frac{\lambda-3\zeta(1-\zeta)}{\zeta^3}\right).
\]
Introducing a new integration variable
\[
  z=\frac{\lambda-3\zeta(1-\zeta)}{\zeta^3},
  \quad
  \frac{dz}{d\zeta}=-\frac{3\lambda}{\zeta^4}
  +\frac{6}{\zeta^3}
  -\frac{3}{\zeta^2},
\]
we obtain
\begin{equation}
  \label{eq:temp_integral}
  a(\lambda)=\int_0^1\frac{a(z)dz}{\frac{3\lambda}{\zeta}-6+3\zeta}.
\end{equation}
Here we used the fact that $a(\zeta)=0$ at $\zeta>1$ and limited the
range of integration accordingly.  At this point we notice that at
$\lambda\to0^+$ the condition $0<z<1$ requires $\zeta/\lambda\to1/3$.
Substituting this result into Eq.~({\ref{eq:temp_integral}}) we find
\begin{equation}
  \label{eq:a_short_small_lambda}
  a(0^+)=\frac13\int_0^1a(z)dz=\frac13,
\end{equation}
where we used the normalization condition (\ref{eq:a_normalization}).
The numerical solution shown in Fig.~\ref{fig:a_lambda_short} is
consistent with Eq.~(\ref{eq:a_short_small_lambda}).

\subsection{B.~~Comparison with the results of Ref.~\cite{imambekov_phenomenology_2009}}
\label{sec:comparison}

The behavior of the spectral function at $\epsilon\to\varepsilon(p)$
can be studied using the approach of
Ref.~\cite{imambekov_phenomenology_2009}.  The latter applies to
nonchiral systems of fermions and, in addition, it assumes Galilean
invariance.  To apply their results to our chiral system, one must
exclude all effects related to the second (left) Fermi point, which in
practice means setting
\begin{equation}
  \label{eq:chiralization}
  V_L=0,
  \quad
  \delta_-=0,
  \quad
  K=1.
\end{equation}
Under these simplifying assumptions, their result is that near the
edge of support the spectral function scales as a power of
$\epsilon-\varepsilon(p)$ with the exponent
\begin{equation}
  \label{eq:mu_general}
  \mu=1-\left(\frac{\delta_+}{2\pi}\right)^2.
\end{equation}
This expression is obtained from Eq.~(14) of
Ref.~\cite{imambekov_phenomenology_2009}, where one should choose
$\{n,\pm\}\to\{0,-\}$ branch of hole excitations.  (The other branches
are affected by the second Fermi point.) The phase shift $\delta_+$
can be obtained by applying the conditions (\ref{eq:chiralization}) to
Eqs.~(8) and (10) of Ref.~\cite{imambekov_phenomenology_2009}, which
do not rely on Galilean invariance,
\begin{equation}
  \label{eq:delta_+}
  \delta_+=2\pi
  \frac{\frac{\partial \varepsilon(k)}{\partial k_F}+v}
  {v_d-v}.
\end{equation}
Here we have replaced the particle density $\rho$ with $k_F/\pi$.  In
the above expression, $\varepsilon(k)$ is the energy of the hole
measured from the Fermi energy which in our case is the energy of the
bosonic excitation $\varepsilon(p)$.  An important difference is that
unlike $k$, our $p$ is measured from the Fermi momentum $k_F$.  Thus
one should substitute into Eq.~(\ref{eq:delta_+}) the energy
$\varepsilon(k)\to\varepsilon(k-k_F)$.  The denominator of
Eq.~(\ref{eq:delta_+}) contains velocity of the hole $v_d$, which in
our case is simply $\varepsilon'(p)=\varepsilon'(k-k_F)$.  As a
result, in the case of strongly interacting chiral fermions considered
in this paper, Eq.~(\ref{eq:delta_+}) yields $\delta_+=-2\pi$.

According to Eq.~(\ref{eq:mu_general}), this value of $\delta_+$
yields power-law scaling of the spectral function with the exponent
$\mu=0$ at $\epsilon\to\varepsilon(p)$.  This limit corresponds to
$\lambda\to0^+$.  Thus, our result (\ref{eq:a_short_small_lambda})
also gives $\mu=0$.

\subsection{C.~~Behavior of $a(\lambda)$ at $\lambda\to1^-$.}
\label{sec:asympt-lambda-1}

The behavior shown in Fig.~\ref{fig:a_lambda_short} suggests that in
the limit $\lambda\to1^-$ the dimensionless spectral function
$a(\lambda)$ approaches zero exponentially.  We will therefore write
it in the form
\begin{equation}
  \label{eq:a_vs_alpha}
  a(\lambda)=e^{-\alpha(\chi)},
  \qquad
  \chi=1-\lambda
\end{equation}
and study the behavior of $\alpha(\chi)$ at $\chi\to0^+$.  We start by
defining the function
\begin{equation}
  \label{eq:Lambda_short}
  z(\zeta)=\frac{\lambda-3\zeta(1-\zeta)}{\zeta^3}
  =\frac{1-\chi-3\zeta(1-\zeta)}{\zeta^3},
\end{equation}
cf.~Eq.~(\ref{eq:integral-equation-short}).  $z(\zeta)$ has a
minimum at $\tilde\zeta=1-\sqrt{\chi}$, near which it behaves as
\begin{equation}
  \label{eq:Lambda_series}
  z(\zeta)\simeq 1-\tilde\chi+3\sqrt{\chi}\,(\zeta-\tilde\zeta)^2,
  \quad
  \tilde\chi=\frac{\chi}{(1-\sqrt\chi)^2}
\end{equation}
where the coefficient of the term $(\zeta-\tilde\zeta)^2$ is written
to leading order in $\chi\ll1$.

We now substitute Eqs.~(\ref{eq:a_vs_alpha}), (\ref{eq:Lambda_short}) and
(\ref{eq:Lambda_series}) into Eq.~(\ref{eq:integral-equation-short}),
\begin{eqnarray*}
  e^{-\alpha(\chi)}
  &\simeq&
  \int_0^1\frac{d\zeta}{\zeta^3}
      e^{-\alpha(\tilde\chi-3\sqrt\chi(\zeta-\tilde\zeta)^2)}\\
  &\simeq&
  e^{-\alpha(\tilde\chi)}
  \int_0^1\frac{d\zeta}{\zeta^3}
      e^{3\alpha'(\tilde\chi)\sqrt\chi(\zeta-\tilde\zeta)^2}
\end{eqnarray*}
We now assume that the derivative $\alpha'(\tilde\chi)$ is negative
and sufficiently large to justify applying the saddle point
approximation to the above integral.  The latter yields
\begin{equation}
  \label{eq:saddle_result}
  e^{\alpha(\tilde\chi)-\alpha(\chi)}
  \simeq
  \sqrt{\frac{\pi}{3|\alpha'(\tilde\chi)|\sqrt\chi}}.
\end{equation}
This approximation is applicable provided that
$|\zeta-\tilde\zeta|\sim1/\sqrt{|\alpha'|\sqrt\chi} \ll 1-\tilde\zeta
= \sqrt\chi$, i.e.,
\begin{equation}
  \label{eq:saddle_condition}
  \chi^{3/2}|\alpha'(\tilde\chi)|\gg1,
\end{equation}
which will be verified later.
To leading order one can expand the exponent in the left-hand side of
Eq.~(\ref{eq:saddle_result}) and obtain
\[
|\alpha'(\tilde\chi)|(\chi-\tilde\chi)
  =\frac12\ln\frac{\pi}{3|\alpha'(\tilde\chi)|\sqrt\chi}
\]
or, substituting $\tilde\chi\simeq\chi+2\chi^{3/2}$ from
Eq.~(\ref{eq:Lambda_series}), 
\begin{equation}
  \label{eq:alpha'equation}
  4\chi^{3/2}|\alpha'(\tilde\chi)|
  =\ln\frac{3|\alpha'(\tilde\chi)|\sqrt\chi}{\pi}.
\end{equation}
We now present $|\alpha'(\tilde\chi)|$ as
\begin{equation}
  \label{eq:alpha'_vs_xi}
  |\alpha'(\tilde\chi)|=\frac{\xi}{4\chi^{3/2}}.
\end{equation}
Equation (\ref{eq:alpha'equation}) then takes the form
\begin{equation}
  \label{eq:xi}
  \xi=\ln\frac{3\xi}{4\pi \chi}.
\end{equation}
To leading order at $\chi\to0^+$, we find
\begin{equation}
  \label{eq:xi_1}
  \xi\simeq\ln\frac{3}{4\pi \chi}.
\end{equation}
Note that this guarantees $\xi\gg1$, i.e., the condition
(\ref{eq:saddle_condition}) for the applicability of the saddle point
approximation (\ref{eq:saddle_result}) is satisfied.  Using
Eqs.~(\ref{eq:alpha'_vs_xi}) and (\ref{eq:xi_1}) and keeping in mind
that $\alpha'<0$, we obtain
\begin{equation}
  \label{eq:alpha_short_1}
  \alpha(\chi)\simeq
  \frac{1}{2\sqrt\chi}\left(\ln\frac{3}{4\pi\chi}-2\right).
\end{equation}
The error in the approximate solution (\ref{eq:xi_1}) of
Eq.~(\ref{eq:xi}) is of the order of $\ln(\ln(1/\chi))$.  Thus the
coefficient in the argument of the logarithm in
Eq.~(\ref{eq:alpha_short_1}) must be omitted.  This yields
\begin{equation}
  \label{eq:alpha_short_main}
  \alpha(\chi)\simeq
  \frac{1}{2\sqrt\chi}\ln\frac{1}{\chi},
\end{equation}
which is equivalent to Eq.~(\ref{eq:a-near-1}) of the main text.

\section{2.~~Asymptotic properties of the spectral function in the
  case of Coulomb interactions}

Here we study the asymptotic properties of the solution of the
integral equation (\ref{eq:integral-equation-Coulomb}) at
$\lambda\to0^+$ and $\lambda\to\infty$.  For convenience, we rewrite
Eq.~(\ref{eq:integral-equation-Coulomb}) as
\begin{equation}
  \label{eq:integral_equation}
  a(\lambda)=\int_0^1\frac{d\zeta}{\zeta}
  a\left(\frac{\lambda-f(\zeta)}{\zeta}\right),
\end{equation}
where
\begin{equation}
  \label{eq:f}
  f(\zeta)=\zeta\ln\frac{1}{\zeta}
  +(1-\zeta)\ln\frac{1}{1-\zeta}.
\end{equation}
We are interested in the solution that satisfies the normalization
condition
\begin{equation}
  \label{eq:a_normalization_Coulomb}
  \int_0^\infty a(z)dz=1
\end{equation}
and vanishes at $\lambda<0$.

\subsection{A.~~Behavior of $a(\lambda)$ at $\lambda\to0^+$}
\label{sec:zero_lambda}

Because $a(\lambda)=0$ at negative $\lambda$, under the conditions
$0<\lambda<{\rm max} f(\zeta)=\ln2$ one can rewrite
Eq.~(\ref{eq:integral_equation}) as
\[
  a(\lambda)=\int_0^\eta\frac{d\zeta}{\zeta}
  a\left(\frac{\lambda-f(\zeta)}{\zeta}\right)
  +\int_{1-\eta}^1\frac{d\zeta}{\zeta}
  a\left(\frac{\lambda-f(\zeta)}{\zeta}\right),
\]
where $\eta$ is defined by
\begin{equation}
  \label{eq:eta_definition}
  f(\eta)=\lambda,
  \quad
  0<\eta<\frac12.
\end{equation}
We are interested in the regime
$\lambda\ll1$, which corresponds to $\eta\ll1$.  In this case the
second integral, in which $\zeta$ is close to 1, is much smaller than
the first one, where $\zeta\leq\eta$.  Omitting it and replacing
$\zeta=(1-u)\eta$, we obtain
\begin{equation}
  \label{eq:intermediate}
  a(\lambda)=\int_0^1\frac{du}{1-u}
  a\left(\frac{f(\eta)-f((1-u)\eta)}{(1-u)\eta}\right).  
\end{equation}
At small $\zeta$ one can approximate
\begin{equation}
  \label{eq:f_approx}
  f(\zeta)\simeq \zeta\ln\frac{1}{\zeta}
\end{equation}
and obtain
\[
  f(\eta)-f((1-u)\eta)\simeq u\eta \ln\frac{1}{\eta}
  -\eta(1-u)\ln\frac{1}{1-u}
  \simeq u\eta \ln\frac{1}{\eta}.
\]
The argument of the function $a(\lambda)$ in the integrand of
Eq.~(\ref{eq:intermediate}) is then
\[
  \frac{f(\eta)-f((1-u)\eta)}{(1-u)\eta}
  \simeq
  \frac{u \ln\frac{1}{\eta}}{1-u}.
\]
Since $a(\lambda)$ falls off rapidly at $\lambda\gg1$, the integral
(\ref{eq:intermediate}) is dominated by the values of $u\sim
1/\ln\frac{1}{\eta}\ll1$.  This enables us to approximate
$1-u\simeq1$ and to extend the integration over $u$ to infinity,
\begin{equation}
  \label{eq:intermediate2}
  a(\lambda)\simeq\int_0^\infty du\,
  a\left(u\ln\frac{1}{\eta}\right).  
\end{equation}
Using the normalization condition (\ref{eq:a_normalization_Coulomb}),
we obtain
\begin{equation}
  \label{eq:a_small_lambda}
  a(\lambda)\simeq\frac{1}{\ln\frac{1}{\eta}},
\end{equation}
where $\eta$ should be obtained by solving the equation
\begin{equation}
  \label{eq:eta_equation}
  \eta\ln\frac{1}{\eta}=\lambda,
\end{equation}
obtained from Eqs.~(\ref{eq:eta_definition}) and (\ref{eq:f_approx}).
Taking the logarithm of both sides of Eq.~(\ref{eq:eta_equation}), we
obtain
\begin{equation}
  \label{eq:eta_equation2}
  \ln\frac{1}{\eta}-\ln\left(\ln\frac{1}{\eta}\right)=\ln\frac1\lambda,
\end{equation}
At very small $\eta$ one can neglect the second term in the left-hand
side of Eq.~(\ref{eq:eta_equation2}) and obtain
\begin{equation}
  \label{eq:a_small_lambda_result1}
  a(\lambda)\simeq\frac{1}{\ln\frac{1}{\lambda}},
\end{equation}
cf. Eq.~(\ref{eq:a_small_lambda_Coulomb}) of the main text.
A more accurate result
\begin{equation}
  \label{eq:a_small_lambda_result2}
  a(\lambda)\simeq\frac{1}{\ln\frac{1}{\lambda}+\ln\big(\ln\frac{1}{\lambda}\big)}.
\end{equation}
is obtained by iterating Eq.~(\ref{eq:eta_equation2}) once.
%\\[1ex]

\subsection{B.~~Behavior of $a(\lambda)$ at $\lambda\to\infty$}
\label{sec:infinite_lambda}

Figure \ref{fig:a_lambda-Coulomb} suggests that at $\lambda\to\infty$
the spectral function falls off exponentially.  We thus present it in
the form
\begin{equation}
  \label{eq:a_vs_beta}
  a(\lambda)=e^{-\beta(\lambda)}
\end{equation}
where the function $\beta(\lambda)$ is expected to approach infinity
at large $\lambda$.  We now examine the integral equation
(\ref{eq:integral_equation}) to obtain asymptotic behavior of
$\beta(\lambda)$ at $\lambda\to\infty$.

To this end we introduce the function
\begin{equation}
  \label{eq:Lambda_Coulomb}
  \Lambda(\zeta)=\frac{\lambda-f(\zeta)}{\zeta}.
\end{equation}
The latter has a minimum at $\tilde\zeta=1-e^{-\lambda}$, near which it
behaves as
\begin{equation}
  \label{eq:Lambda_Coulomb_series}
  \Lambda(\zeta)\simeq
  \tilde\lambda
  +\frac{e^\lambda}{2\tilde\zeta^2}(\zeta-\tilde\zeta)^2,
  \qquad
  \tilde\lambda=\lambda+\ln(1-e^{-\lambda}).
\end{equation}
Assuming that not only $\beta(\lambda)$, but also its derivative is
large at $\lambda\to\infty$, the integral in
Eq.~(\ref{eq:integral_equation}) can be evaluated using the saddle
point approximation:
\begin{eqnarray}
  e^{-\beta(\lambda)}
  &\simeq&
  e^{-\beta(\tilde\lambda)}
  \int_0^1\frac{d\zeta}{\zeta}
  e^{-\beta'(\tilde\lambda)\frac{e^\lambda}{2\tilde\zeta^2}(\zeta-\tilde\zeta)^2}
\nonumber\\
  &\simeq&
  e^{-\beta(\tilde\lambda)}\sqrt{\frac{2\pi}{\beta'(\tilde\lambda)e^{\lambda}}}.
  \label{eq:intermediate3}
\end{eqnarray}
The condition of applicability of the saddle point approximation is
that the typical $|\zeta-\tilde\zeta|$ in the integral is small
compared to $1-\tilde\zeta$, i.e.,
\[
   \frac{1}{\sqrt{\beta'(\tilde\lambda)e^{\lambda}}}\ll e^{-\lambda},
\]
is satisfied for our result (\ref{eq:beta'result}).
Rewriting Eq.~(\ref{eq:intermediate3}) as
\[
  e^{\beta(\tilde\lambda)-\beta(\lambda)}\simeq
  \sqrt{\frac{2\pi}{\beta'(\tilde\lambda)e^{-\lambda}}}\,
  e^{-\lambda}
\]
and expanding $\beta(\tilde\lambda)-\beta(\lambda)\simeq
\beta'(\tilde\lambda) (\tilde\lambda-\lambda)$ we obtain
\begin{equation}
  \label{eq:beta'eq}
  \beta'(\tilde\lambda)e^{-\lambda}
  \simeq\lambda
  +\frac12\ln\frac{\beta'(\tilde\lambda)e^{-\lambda}}{2\pi}.
\end{equation}
To leading order at $\lambda\to\infty$ one can replace
$\tilde\lambda\to\lambda$ and neglect the second term on the
right-hand side.  This yields
\begin{equation}
  \label{eq:beta'result}
  \beta'(\lambda)\simeq\lambda e^\lambda
\end{equation}
and thus
\begin{equation}
  \label{eq:beta_result}
  \beta(\lambda)\simeq\lambda e^\lambda.
\end{equation}
Note that we have replaced $(\lambda-1)\to\lambda$ in the prefactor,
because the leading order correction in Eq.~(\ref{eq:beta'result}) is
$\frac12(\ln\lambda)e^\lambda\gg e^\lambda$.  Equation
(\ref{eq:beta_result}) is equivalent to $\ln a(\lambda)\simeq -\lambda
e^\lambda$, as quoted in the main text.


\begin{thebibliography}{14}%
\makeatletter
\providecommand \@ifxundefined [1]{%
 \@ifx{#1\undefined}
}%
\providecommand \@ifnum [1]{%
 \ifnum #1\expandafter \@firstoftwo
 \else \expandafter \@secondoftwo
 \fi
}%
\providecommand \@ifx [1]{%
 \ifx #1\expandafter \@firstoftwo
 \else \expandafter \@secondoftwo
 \fi
}%
\providecommand \natexlab [1]{#1}%
\providecommand \enquote  [1]{``#1''}%
\providecommand \bibnamefont  [1]{#1}%
\providecommand \bibfnamefont [1]{#1}%
\providecommand \citenamefont [1]{#1}%
\providecommand \href@noop [0]{\@secondoftwo}%
\providecommand \href [0]{\begingroup \@sanitize@url \@href}%
\providecommand \@href[1]{\@@startlink{#1}\@@href}%
\providecommand \@@href[1]{\endgroup#1\@@endlink}%
\providecommand \@sanitize@url [0]{\catcode `\\12\catcode `\$12\catcode
  `\&12\catcode `\#12\catcode `\^12\catcode `\_12\catcode `\%12\relax}%
\providecommand \@@startlink[1]{}%
\providecommand \@@endlink[0]{}%
\providecommand \url  [0]{\begingroup\@sanitize@url \@url }%
\providecommand \@url [1]{\endgroup\@href {#1}{\urlprefix }}%
\providecommand \urlprefix  [0]{URL }%
\providecommand \Eprint [0]{\href }%
\providecommand \doibase [0]{https://doi.org/}%
\providecommand \selectlanguage [0]{\@gobble}%
\providecommand \bibinfo  [0]{\@secondoftwo}%
\providecommand \bibfield  [0]{\@secondoftwo}%
\providecommand \translation [1]{[#1]}%
\providecommand \BibitemOpen [0]{}%
\providecommand \bibitemStop [0]{}%
\providecommand \bibitemNoStop [0]{.\EOS\space}%
\providecommand \EOS [0]{\spacefactor3000\relax}%
\providecommand \BibitemShut  [1]{\csname bibitem#1\endcsname}%
\let\auto@bib@innerbib\@empty
%</preamble>
\bibitem [{\citenamefont {Haldane}(1981)}]{haldane_luttinger_1981}%
  \BibitemOpen
  \bibfield  {author} {\bibinfo {author} {\bibfnamefont {F.~D.~M.}\
  \bibnamefont {Haldane}},\ }\bibfield  {title} {\bibinfo {title} {'{Luttinger}
  liquid theory' of one-dimensional quantum fluids. {I}. {Properties} of the
  {Luttinger} model and their extension to the general {1D} interacting
  spinless {Fermi} gas},\ }\href {https://doi.org/10.1088/0022-3719/14/19/010}
  {\bibfield  {journal} {\bibinfo  {journal} {J. Phys. C: Solid State Phys.}\
  }\textbf {\bibinfo {volume} {14}},\ \bibinfo {pages} {2585} (\bibinfo {year}
  {1981})}\BibitemShut {NoStop}%
\bibitem [{\citenamefont {Giamarchi}(2004)}]{giamarchi_quantum_2004}%
  \BibitemOpen
  \bibfield  {author} {\bibinfo {author} {\bibfnamefont {T.}~\bibnamefont
  {Giamarchi}},\ }\href@noop {} {\emph {\bibinfo {title} {Quantum physics in
  one dimension}}}\ (\bibinfo  {publisher} {Clarendon},\ \bibinfo {address}
  {Oxford},\ \bibinfo {year} {2004})\BibitemShut {NoStop}%
\bibitem [{\citenamefont {Kane}\ and\ \citenamefont
  {Fisher}(1992)}]{kane_transmission_1992}%
  \BibitemOpen
  \bibfield  {author} {\bibinfo {author} {\bibfnamefont {C.~L.}\ \bibnamefont
  {Kane}}\ and\ \bibinfo {author} {\bibfnamefont {M.~P.~A.}\ \bibnamefont
  {Fisher}},\ }\bibfield  {title} {\bibinfo {title} {Transmission through
  barriers and resonant tunneling in an interacting one-dimensional electron
  gas},\ }\href {https://doi.org/10.1103/PhysRevB.46.15233} {\bibfield
  {journal} {\bibinfo  {journal} {Phys. Rev. B}\ }\textbf {\bibinfo {volume}
  {46}},\ \bibinfo {pages} {15233} (\bibinfo {year} {1992})}\BibitemShut
  {NoStop}%
\bibitem [{\citenamefont {Furusaki}\ and\ \citenamefont
  {Nagaosa}(1993)}]{furusaki_single-barrier_1993}%
  \BibitemOpen
  \bibfield  {author} {\bibinfo {author} {\bibfnamefont {A.}~\bibnamefont
  {Furusaki}}\ and\ \bibinfo {author} {\bibfnamefont {N.}~\bibnamefont
  {Nagaosa}},\ }\bibfield  {title} {\bibinfo {title} {Single-barrier problem
  and {Anderson} localization in a one-dimensional interacting electron
  system},\ }\href {https://doi.org/10.1103/PhysRevB.47.4631} {\bibfield
  {journal} {\bibinfo  {journal} {Phys. Rev. B}\ }\textbf {\bibinfo {volume}
  {47}},\ \bibinfo {pages} {4631} (\bibinfo {year} {1993})}\BibitemShut
  {NoStop}%
\bibitem [{\citenamefont {Wen}(1992)}]{wen_theory_1992}%
  \BibitemOpen
  \bibfield  {author} {\bibinfo {author} {\bibfnamefont {X.-G.}\ \bibnamefont
  {Wen}},\ }\bibfield  {title} {\bibinfo {title} {Theory of the edge states in
  fractional quantum {Hall} effects},\ }\href
  {https://doi.org/10.1142/S0217979292000840} {\bibfield  {journal} {\bibinfo
  {journal} {Int. J. Mod. Phys. B}\ }\textbf {\bibinfo {volume} {06}},\
  \bibinfo {pages} {1711} (\bibinfo {year} {1992})}\BibitemShut {NoStop}%
\bibitem [{\citenamefont {Chang}(2003)}]{chang_chiral_2003}%
  \BibitemOpen
  \bibfield  {author} {\bibinfo {author} {\bibfnamefont {A.~M.}\ \bibnamefont
  {Chang}},\ }\bibfield  {title} {\bibinfo {title} {Chiral {Luttinger} liquids
  at the fractional quantum {Hall} edge},\ }\href
  {https://doi.org/10.1103/RevModPhys.75.1449} {\bibfield  {journal} {\bibinfo
  {journal} {Rev. Mod. Phys.}\ }\textbf {\bibinfo {volume} {75}},\ \bibinfo
  {pages} {1449} (\bibinfo {year} {2003})}\BibitemShut {NoStop}%
\bibitem [{\citenamefont {Halperin}(1982)}]{halperin_quantized_1982}%
  \BibitemOpen
  \bibfield  {author} {\bibinfo {author} {\bibfnamefont {B.~I.}\ \bibnamefont
  {Halperin}},\ }\bibfield  {title} {\bibinfo {title} {Quantized {Hall}
  conductance, current-carrying edge states, and the existence of extended
  states in a two-dimensional disordered potential},\ }\href
  {https://doi.org/10.1103/PhysRevB.25.2185} {\bibfield  {journal} {\bibinfo
  {journal} {Phys. Rev. B}\ }\textbf {\bibinfo {volume} {25}},\ \bibinfo
  {pages} {2185} (\bibinfo {year} {1982})}\BibitemShut {NoStop}%
\bibitem [{\citenamefont {Martin}\ and\ \citenamefont
  {Matveev}()}]{2109.06220}%
  \BibitemOpen
  \bibfield  {author} {\bibinfo {author} {\bibfnamefont {I.}~\bibnamefont
  {Martin}}\ and\ \bibinfo {author} {\bibfnamefont {K.~A.}\ \bibnamefont
  {Matveev}},\ }\href@noop {} {\bibinfo {title} {Scar states in a system of
  interacting chiral fermions}},\ \Eprint
  {https://arxiv.org/abs/arXiv:2109.06220v1} {arXiv:2109.06220v1} \BibitemShut
  {NoStop}%
\bibitem [{\citenamefont {Imambekov}\ and\ \citenamefont
  {Glazman}(2009)}]{imambekov_phenomenology_2009}%
  \BibitemOpen
  \bibfield  {author} {\bibinfo {author} {\bibfnamefont {A.}~\bibnamefont
  {Imambekov}}\ and\ \bibinfo {author} {\bibfnamefont {L.~I.}\ \bibnamefont
  {Glazman}},\ }\bibfield  {title} {\bibinfo {title} {Phenomenology of
  {One}-{Dimensional} {Quantum} {Liquids} {Beyond} the {Low}-{Energy}
  {Limit}},\ }\href {https://doi.org/10.1103/PhysRevLett.102.126405} {\bibfield
   {journal} {\bibinfo  {journal} {Phys. Rev. Lett.}\ }\textbf {\bibinfo
  {volume} {102}},\ \bibinfo {pages} {126405} (\bibinfo {year}
  {2009})}\BibitemShut {NoStop}%
\bibitem [{\citenamefont {Iordanskii}\ and\ \citenamefont
  {Pitaevskii}(1978)}]{iordanskii_properties_1978}%
  \BibitemOpen
  \bibfield  {author} {\bibinfo {author} {\bibfnamefont {S.~V.}\ \bibnamefont
  {Iordanskii}}\ and\ \bibinfo {author} {\bibfnamefont {L.~P.}\ \bibnamefont
  {Pitaevskii}},\ }\bibfield  {title} {\bibinfo {title} {Properties of the
  endpoint of a multiphonon spectrum},\ }\href@noop {} {\bibfield  {journal}
  {\bibinfo  {journal} {JETP Lett.}\ }\textbf {\bibinfo {volume} {27}},\
  \bibinfo {pages} {621} (\bibinfo {year} {1978})}\BibitemShut {NoStop}%
\bibitem [{\citenamefont {Khoze}\ and\ \citenamefont
  {Reiness}(2019)}]{khoze_review_2019}%
  \BibitemOpen
  \bibfield  {author} {\bibinfo {author} {\bibfnamefont {V.~V.}\ \bibnamefont
  {Khoze}}\ and\ \bibinfo {author} {\bibfnamefont {J.}~\bibnamefont
  {Reiness}},\ }\bibfield  {title} {\bibinfo {title} {Review of the
  semiclassical formalism for multiparticle production at high energies},\
  }\href {https://doi.org/10.1016/j.physrep.2019.06.004} {\bibfield  {journal}
  {\bibinfo  {journal} {Phys. Rep.}\ }\textbf {\bibinfo {volume} {822}},\
  \bibinfo {pages} {1} (\bibinfo {year} {2019})}\BibitemShut {NoStop}%
\bibitem [{\citenamefont {Khodas}\ \emph {et~al.}(2007)\citenamefont {Khodas},
  \citenamefont {Pustilnik}, \citenamefont {Kamenev},\ and\ \citenamefont
  {Glazman}}]{khodas_fermi-luttinger_2007}%
  \BibitemOpen
  \bibfield  {author} {\bibinfo {author} {\bibfnamefont {M.}~\bibnamefont
  {Khodas}}, \bibinfo {author} {\bibfnamefont {M.}~\bibnamefont {Pustilnik}},
  \bibinfo {author} {\bibfnamefont {A.}~\bibnamefont {Kamenev}},\ and\ \bibinfo
  {author} {\bibfnamefont {L.~I.}\ \bibnamefont {Glazman}},\ }\bibfield
  {title} {\bibinfo {title} {Fermi-{Luttinger} liquid: {Spectral} function of
  interacting one-dimensional fermions},\ }\href
  {https://doi.org/10.1103/PhysRevB.76.155402} {\bibfield  {journal} {\bibinfo
  {journal} {Phys. Rev. B}\ }\textbf {\bibinfo {volume} {76}},\ \bibinfo
  {pages} {155402} (\bibinfo {year} {2007})}\BibitemShut {NoStop}%
\bibitem [{\citenamefont {Hardy}\ and\ \citenamefont
  {Ramanujan}(1918)}]{hardy_asymptotic_1918}%
  \BibitemOpen
  \bibfield  {author} {\bibinfo {author} {\bibfnamefont {G.~H.}\ \bibnamefont
  {Hardy}}\ and\ \bibinfo {author} {\bibfnamefont {S.}~\bibnamefont
  {Ramanujan}},\ }\bibfield  {title} {\bibinfo {title} {Asymptotic {FormulaÃ¦}
  in {Combinatory} {Analysis}},\ }\href
  {https://doi.org/10.1112/plms/s2-17.1.75} {\bibfield  {journal} {\bibinfo
  {journal} {Proc. London Math. Soc.}\ }\textbf {\bibinfo {volume} {s2-17}},\
  \bibinfo {pages} {75} (\bibinfo {year} {1918})}\BibitemShut {NoStop}%
\bibitem [{one()}]{onefootnote}%
  \BibitemOpen
  \href@noop {} {}\bibinfo {note} {For details, see Supplemental
  Material.}\BibitemShut {Stop}%
\end{thebibliography}
\end{document}